\documentclass[journal]{IEEEtran}
\usepackage[utf8]{inputenc}
\usepackage[T1]{fontenc} 
\usepackage[english]{babel}
\usepackage[pdftex]{graphicx}
\usepackage{tikz,pgfplots}
\usepackage{amsmath,amssymb,graphicx}
\usepackage{comment}
\usepackage{pgf,tikz}
\usepgfplotslibrary{units}
\usetikzlibrary{spy,backgrounds}
\usepackage{pgfplotstable}\usetikzlibrary{calc}
\usetikzlibrary{shapes,arrows}
\usetikzlibrary{pgfplots.statistics}
\usepackage{adjustbox}
\usepackage{subcaption}
\usepackage{csquotes}
\usepackage{hyperref}
\usepackage{booktabs}
\usepackage{lipsum}

\usetikzlibrary{shapes,snakes}
\usetikzlibrary{arrows}
    \usepgfplotslibrary{statistics}
    \pgfplotsset{compat=1.8}
    \pgfmathdeclarefunction{fpumod}{2}{%
        \pgfmathfloatdivide{#1}{#2}%
        \pgfmathfloatint{\pgfmathresult}%
        \pgfmathfloatmultiply{\pgfmathresult}{#2}%
        \pgfmathfloatsubtract{#1}{\pgfmathresult}%
        \pgfmathfloatifapproxequalrel{\pgfmathresult}{#2}{\def\pgfmathresult{5}}{}%
    }

\DeclareMathOperator*{\argmax}{arg\,max}

\newcommand{\cev}[1]{\reflectbox{\ensuremath{\vec{\reflectbox{\ensuremath{#1}}}}}}

\newlength\mylength
\newcommand\copyrighttext{%
  \footnotesize \textit{ Preprint posted to arXiv. } Copyright 2020 by the authors.}
\newcommand\copyrightnotice{%
\begin{tikzpicture}[remember picture,overlay]
\node[anchor=south,yshift=10pt] at (current page.south) {\fbox{\parbox{\dimexpr\textwidth-\fboxsep-\fboxrule\relax}{\copyrighttext}}};
\end{tikzpicture}%
}

\begin{document}
%
\title{Segmentation and Optimal Region Selection of Physiological Signals using Deep Neural Networks and Combinatorial Optimization}


\author{\IEEEauthorblockN{Jorge Oliveira$^{1}$,
Margarida Carvalho $^{2}$, Diogo Marcelo Nogueira $^{3}$ and
Miguel Coimbra$^{4}$, \textit{Senior Member, IEEE}.}\\

\thanks{$^{1}$ is with Instituto de Telecomunica\c c\~oes, Faculdade de Ciências da Universidade do Porto, Rua do Campo Alegre 1021/1055, 4169-007 Porto, Portugal.$^{2}$ is with CIRRELT and D\'epartement d'informatique et de recherche op\'erationnelle, Universit\'e de Montr\'eal, Montr\'eal, Canada;$^{3,4}$ are with INESC TEC, Rua Dr. Roberto Frias, 4200-465 Porto, Portugal. This work is a result of the Project DigiScope2 (POCI-01-0145-FEDER-029200 - PTDC/CCI-COM/29200/2017), funded by Fundo Europeu de Desenvolvimento Regional (FEDER), through Programa Operacional Competitividade e Internacionalização (POCI), by national funds, through Funda\c{c}\~ao para a Ci\^encia e Tecnologia (FCT), and partially financed by Institut de valorisation des donn\'ees and Fonds de Recherche du Qu\'ebec, through the FRQ-IVADO Research Chair in Data Science for Combinatorial Game Theory. 

 Corresponding author: J. Oliveira (email: \url{jhs@isep.ipp.pt}).}}


\IEEEtitleabstractindextext{%
\begin{abstract}

 Physiological signals, such as the electrocardiogram and the phonocardiogram are very often corrupted by noisy sources. Usually, artificial intelligent algorithms analyze the signal regardless of its quality. On the other hand, physicians use a completely orthogonal strategy. They do not assess the entire recording, instead they search for a segment where the fundamental and abnormal waves are easily detected, and only then a prognostic is attempted.
 

 Inspired by this fact, a new algorithm that automatically selects an optimal segment for a post-processing stage, according to a criteria defined by the user is proposed. In the process, a Neural Network is used to compute the output state probability distribution for each sample. Using the aforementioned quantities, a graph is designed, whereas state transition constraints are physically imposed into the graph and a set of constraints are used to retrieve a subset of the recording that maximizes the likelihood function, proposed by the user.


 The developed framework is tested and validated in two applications. In both cases, the system performance is boosted significantly, e.g in heart sound segmentation, sensitivity increases $2.4\%$ when compared to the standard approaches in the literature. 
\end{abstract}

}

\maketitle
\copyrightnotice

\IEEEdisplaynontitleabstractindextext

\IEEEpeerreviewmaketitle

\section{Introduction}

The accurate interpretation of physiological signals, such as the electrocardiogram (ECG) and the phonocardiogram (PCG) is a very demanding task. A medical student needs to listen around 500 repetitions of each type of murmur in order to learn how to identify them properly \cite{BARRETT_2004}. Furthermore, although medical students are capable of interpreting the primary ECG parameters, their ability to recognize ECG signs of emergencies and common heart abnormalities is low \cite{Kopec:2015}. Only 58\% of the students are able to recognize common ECG abnormalities such as ischemia, rhythm disorder, and cardiac chambers hypertrophy \cite{Kopec:2015}. Therefore, the computerized interpretation of physiological signals can perhaps reduce interpretation errors, especially in places where trained readers are not available \cite{Harold:2019}. On the other hand, these signals are often corrupted by distinct kind of artifacts and noisy sources, e.g. instrumentation noise in ECG signals, body sound noises in a PCG signal, head movement noise in an electroencephalogram signal. Noisy samples can lead to an increase of false alarms or limit the capability of algorithms to detect abnormal waves. Mostly of the pre-processing steps use denoising techniques, such as filtering to suppress or attenuate noisy components or artifacts in physiological signals \cite{Strang:1996}. But, the changes in the waveform made by filters with a narrow band pass frequency, can lead accidentally to a wrong diagnose, e.g. ECG arrhythmia \cite{Satija:2019}. In PCG signals, removing artifacts and noisy components from the signal, results in losses of information since the frequency content of artifacts and heart sound waves usually do overlap \cite{Kumar:2014}.
In order to address this problem, some algorithms first attempt to access automatically the quality of the recording by grading it into quality groups such as acceptable/unacceptable, acceptable/intermediate/unacceptable, excellent/very good/good/bad \cite{Maier:2016}. These categories are usually based on signal quality index values \cite{Morgado:2015}. 
In general physiological signals with a low grade are discarded and not further processed, regardless of their information content \cite{Satija:2019}. These signals are often seen in real-world scenarios, namely in acoustic signals such as the PCG signal, where sometimes controllable conditions are simply not possible, e.g. to perform a heart sound auscultation during an emergency situation. Thus, a robust system which is capable of operating in very \enquote{aggressive} noisy conditions and still be able to retrieve useful information to the clinician is needed.

In this paper, the aforementioned problem is addressed. Our algorithm does not attempt to segment the entire signal, but instead it looks for a continuous fixed-length window where the likelihood function defined by the user is optimized. Inside of such a window, the algorithm attempts to decode \enquote{true} state sequence of events, although restricted to the physiological state transition constraints, which are inherent to each signal. 
From the application point-of-view, a physician can observe of what the algorithm considers to be the most suitable segment of the recording for a further posterior analysis. In the process, noisy and undesired segments are automatically removed without the physician perception. Furthermore using the proposed algorithm, the physician decides the recording duration to be retrieved by the system, thus adjusting to his own needs.  We believe that by using our proposed system, the analysis of physiological signals can become simpler and more tractable to physicians, and hopefully support and help them to take better clinical decisions.

\subsection{Contribution}

Our main contributions are:
\begin{itemize}
    \item A new method to select the optimal region of interest for a further post-processing stage, according to a criterion defined by the user.
    \item A new method to impose physiological constraints to the output of a Neural Network (NN) algorithm, by solving a graph optimization problem.
\end{itemize}

This paper is organized as follow: in Section~\ref{Output_Probability_Distribution} the computation of the state output probability distribution is explained. In Section~\ref{G_otimization}, state output probability distributions and physiological constraints are embedded in different optimization problems for which several out-of-the-box solvers exist. In Sections~\ref{PCG_CASE_STUDY} and~\ref{ECG_CASE_STUDY}, two case studies are presented to the reader, the segmentation of PCG and ECG signals, respectively. Finally, in Section~\ref{sec:Conclusions}, conclusions are drawn.

\section{Computing State Output Probability Distributions}\label{Output_Probability_Distribution}

From the vast possibilities of NN architectures, in this paper, a bidirectional long-short term memory (LSTM) network is used in order to compute the conditional state output probability distributions. Although, CNNs and MLPs are also valid options, our choice is based on the fact that LSTM models are capable of tracking long-term dependencies in the time series \cite{Greff:2017}. Furthermore, LSTM models have recently succeed in detecting abnormal waves in PCG and ECG signals \cite{Latif:2018} \cite{Chauhan:2015} respectively.

\subsection{LSTM model}
\begin{figure}[t]
  \centering
  \includegraphics[width=0.50\textwidth]{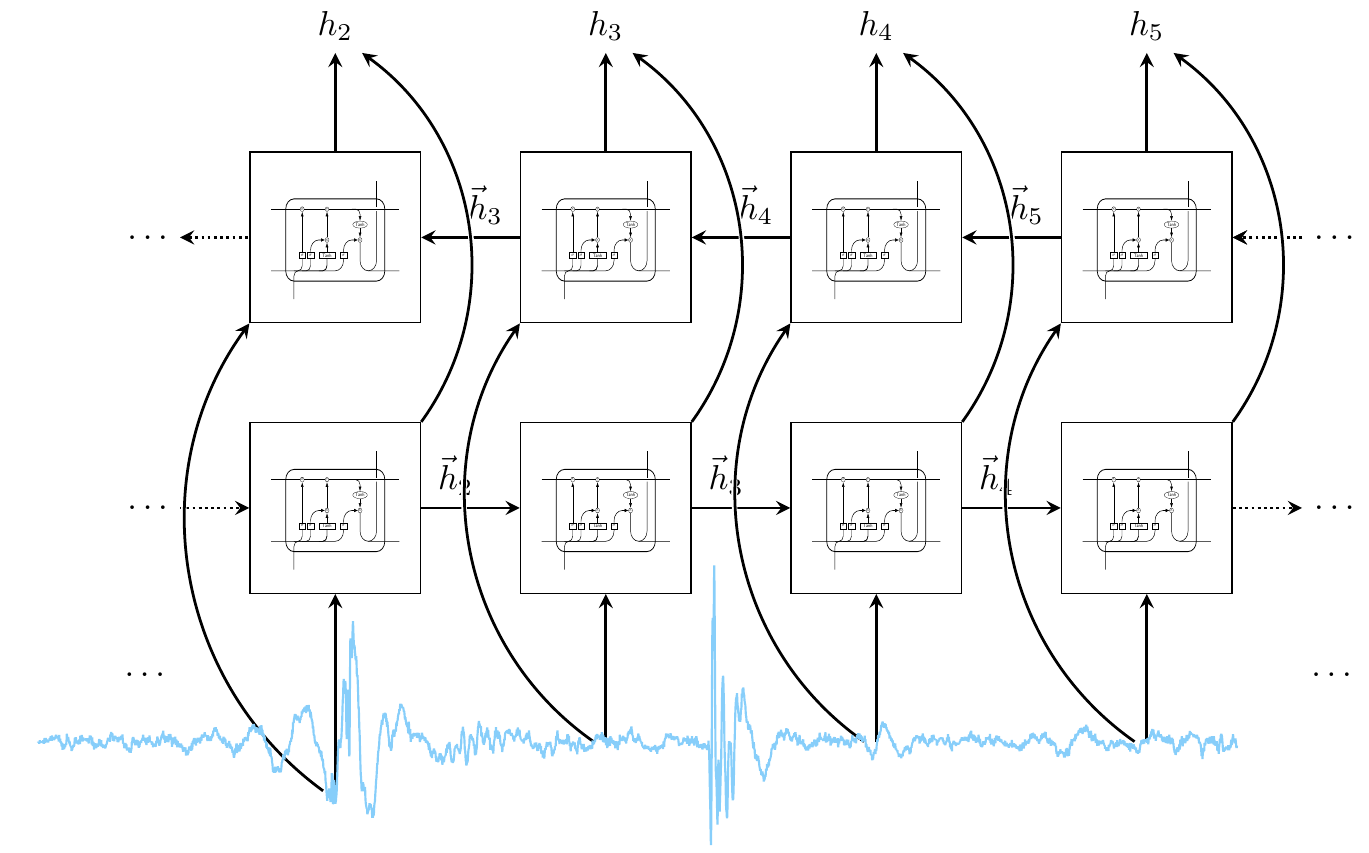}
  \caption{A scheme of the bidirectional LSTM architecture implemented in this work. Two layers of LSTM cells are paired together, adapted from \cite{Greff:2017}.}
  \label{fig:bidirectional_lstm}
\end{figure}

 A LSTM network is a specific type of recurrent neural network (RNN), designed mainly to address the problem of vanishing gradient \cite{Greff:2017}. Given a sequence of feature vectors $\textbf{X} = (x_1, \dots, x_T )$ of length $T$, a standard LSTM network processes sequentially each input feature vector $x$ and generates a sequence of hidden state vectors $H = (h_1, \dots, h_T)$. In our current model two layers are paired together. In the lower layer, information flows forward, from time instant $t=1$ to $t=T$. As a result, the hidden state ($\Vec{h_t} \, \in \mathbb{R}^{M \times 1}$) and the cell state ($\Vec{c_t} \, \in \mathbb{R}^{M \times 1}$) vectors at time $t$ are dependent on past hidden and cell state vectors respectively, $(\Vec{h}_k,\Vec{c}_k) \, \forall k \in \,\{1, \dots, t-1\}$. In the upper layer,  information flows backward, from time instant $t=T$ to $t=1$. As a result, the hidden state ($\cev{h_t} \, \in \mathbb{R}^{M \times 1}$) and the cell state ($\cev{c_t} \, \in \mathbb{R}^{M \times 1}$) vectors at time $t$  are dependent on future hidden and cell state vectors respectively, $(\cev{h}_k,\cev{c}_k) \, \forall k \in \,\{T, \dots, t+1\}$, see Figure \ref{fig:bidirectional_lstm}. Finally, hidden state vectors ($\Vec{h_t} \,, \cev{h_t}$) associated to the same time instant t are merged to form  $h_t \in \mathbb{R}^{2M \times 1} $, see Figure \ref{fig:bidirectional_lstm}. The final matrix $h \in \mathbb{R}^{2M \times T}$ is saved for a further processing. 
 In order to compute $\Vec{h}_t$ and $\Vec{c}_t$, the following equations are implemented in each cell node in the lower layer of the network:
 
 \begin{align}
 \Vec{i}_t = \tanh(\Vec{W}_{xi} x_t + \Vec{W}_{hi}\Vec{h}_{t-1} + \Vec{b}_i) \\
 \Vec{f}_t = \sigma(\Vec{W}_{xf} x_t + \Vec{W}_{hf}\Vec{h}_{t-1} + \Vec{b}_f)\\
 \Vec{o}_t = \tanh(\Vec{W}_{xo} x_t + \Vec{W}_{ho}\Vec{h}_{t-1} + \Vec{b}_o) \\
 \Vec{c}_t = \Vec{c}_{t-1} \odot \Vec{f}_t + \Vec{i}_t \odot \sigma(\Vec{W}_{xj} x_t + \Vec{W}_{hj}\Vec{h}_{t-1} + \Vec{b}_j) \\
 \Vec{h}_t = \tanh(\Vec{c}_t) \odot \Vec{o}_t
 \end{align} 
 
 In the above equations, $\pmb{W_{x \cdot}} \in \Re^{M \times N}$ denotes the weight input matrices,  $\pmb{W_{h \cdot}} \in \Re^{M \times M}$ denotes the weight hidden matrices and $b \in \Re^{M \times 1}$ the bias vectors. The $\Vec{i}, \, \Vec{f},\, \Vec{o}  \in \Re^{M \times 1}$ represents the forward input, forget and output gate respectively. The $\odot$ is a element-wise vector product, $\sigma$ denotes a softmax activation function, $\tanh$ denotes hyperbolic tangent activation function. Furthermore, $N$ corresponds to the number of features extracted by the system and $M$ is the memory size of each cell node. Note that similar equations also exist for $\cev{h_t}$ and $ \cev{c_t}$ respectively.
 

After each input feature vector is processed, the matrix $h$ is going to be fed into a MLP, column by column. In this paper, the MLP does not have any hidden layer and the output layer is fixed to the size $L$, where $L$ is the number of output states. In the output layer, the softmax activation function is used in order to compute the state output probability distribution at time $t$, $p_t = \sigma(\pmb{W}_{out} h_t)$, where $\pmb{W}_{out} \in \Re^{L \times 2M}$ is the weight output matrix. 

\subsection{Standard Approaches}\label{Standard_Approaches}
In order to compute the most likely hidden state at time $t$, the standard approaches usually apply an argmax function:
\begin{equation}
\argmax_k \, [P(p_{t,k} | \pmb{X} = (x_1 , \dots, x_T))],
\label{argmax_k}
\end{equation}
for all $p_{t,k} \in S =  \lbrace 0, \dots, L-1 \rbrace$. The major limitations of this approach are: it is not guaranteed that the system returns an acceptable state sequence of events neither if it is the most likely one.



\section{Global Optimization}\label{G_otimization}

The neural network explained in the previous section, allow us to predict the most likely hidden state for each sample of the signal. However, a non-acceptable solution might be determined, if one simply picks the most likely state for each input sample, determined by the state output probability distribution of the NN. If at time $t$ the hidden state is $s$, then at $t+1$, the only possible states are $s$ and $s+1 \pmod  m$. For sake of simplicity, in the remaining of the paper we drop the module operator. Therefore, our goal is to find the most likely hidden state sequence of events. This can be stated as an integer quadratically constrained optimization problem:
\begin{subequations}
\begin{alignat}{5}
a^*=&\argmax_a  &\sum_{t=1}^{T}\sum_{s \in S} p_{ts} a_{ts}\label{Obj}\\
& \ \ s.t. & \sum_{s \in S} a_{ts} =1  &   \quad \forall t =1, \ldots,T \label{const1}\\
&      & \left(a_{t-1}\right)^T \textbf{Q} a_t = 1 & \quad \forall t =2,\ldots,T \label{const2}\\
& & a_{ts} \in \lbrace 0,1 \rbrace & \quad \forall t =1, \ldots,T,\forall s \in S \label{const3}
\end{alignat}
\label{QuadraticOPT}
\end{subequations}
where 
\begin{equation*} \textbf{Q}=
\left( \begin{matrix}
1 & 1 & 0 &  0 & \cdots       & 0\\
0 & 1 & 1 & 0  & \cdots       & 0 \\
\vdots &   &      &    \ddots     &  & \vdots\\
1 & 0 & 0 & \cdots & 0 & 1
 \end{matrix} \right),
 a_t = \left( \begin{matrix}
a_{t0}\\
a_{t1}\\
\vdots \\
a_{t,L-1}
 \end{matrix} \right). 
 \end{equation*}
The binary decision variables $a_{ts}$ take value 1 if at time $t$ the corresponding hidden state is $s$, and 0 otherwise. The objective function~\eqref{Obj}  represents the likelihood of a state sequence. By maximizing the objective function~\eqref{Obj}, we are determining the sequence $a^*$ for which its likelihood is maximized. Note that other optimization criteria can be used in the objective function providing this formulation with flexibility.
Constraints~\eqref{const1} imply that for each sample, exactly one hidden state must be selected. Constraints~\eqref{const2} enforce physiological state transitions. Constraints~\eqref{const3} restrict the decision variables to binary values. The product of binary variables in Constraints~\eqref{const2} can be linearized in a standard way. If $a_{ts}$ and $a_{t',s'}$ are binary variables, then the product $a_{ts} \cdot a_{t',s'}$ can be equivalently replaced by a new continuous variable $z$ with the following additional constraints: $z\leq a_{ts}$, $z \leq a_{t',s'}$ and $z \geq a_{ts}+a_{t',s'}-1$. In this way, we obtain an integer linear program that triplicated the number of variables of Problem~\eqref{QuadraticOPT}, but hopefully simpler to solve, since it is linear. 

\subsection{Optimal Region Selection}
In general, signals are noisy and therefore, we might aim to find a fixed time window were our prediction of the hidden state sequence is probabilistically  more accurate. In other words, the goal is to determine the best $J$ seconds of our prediction. To that end, we modify Problem~\eqref{QuadraticOPT} as follows:
\begin{subequations}
\begin{alignat}{5}
&\max_{a,b}  &\sum_{t=1}^{T}\sum_{s \in S} p_{ts} a_{ts}\label{Obj1}\\
& s.t. & \sum_{t=1}^T \sum_{s \in S} a_{ts} =J F \label{const4}\\
& & \sum_{t=2}^T \left(b_{t-1}-b_t\right)^2 =2 \label{const5}\\
& & b_t+\sum_{s \in S} a_{ts} = 1 &\qquad \forall t =1, \ldots,T \label{const6}\\
&      & b_t+a_{t-1}^T \textbf{Q} a_t = 1 & \qquad \forall t =2,\ldots,T \label{const7}\\
& & a_{ts} \in \lbrace 0,1 \rbrace & \qquad \forall t =1, \ldots,T , \forall s \in S \label{const8}\\
& & b_t \in \lbrace 0,1 \rbrace &\qquad \forall t =1,\ldots, T\label{const9}
\end{alignat}
\label{ExtraConstraints}
\end{subequations}
where the parameter $F$ is the sample frequency and $b_t$ are new additional decision variables that take value 1 if no state is assigned at time $t$, and 0 otherwise. Constraint~\eqref{const4} enforces that state assignment is only performed for $J$ seconds which together with Constraint~\eqref{const5}  enforces that these $J$ seconds are consecutive. For consecutive segments with no state assignment the sum of $\left(b_{t-1}-b_t\right)^2$ is 0. When in a time $t'$, states start to be assigned, $\left(b_{t'-1}-b_{t'}\right)^2$ is 1. Analogously, when in a time $t''$, stop being assigned, $\left(b_{t''-1}-b_{t''}\right)^2$ is 1. By the model construction, note that the best $J$ seconds cannot start in time 1 or finish in time $T$. Given the long size of the signals, we can admit that this is not a strong limitation of the model. Nevertheless, in Section~\ref{app:graph}, an alternative formulation is proposed which overcomes this limitation. Finally, Constraints~\eqref{const6} and~\eqref{const7} are an adaptation of Constraints~\eqref{const1} and~\eqref{const2}.

Analogously to Problem~\eqref{QuadraticOPT}, Problem~\eqref{ExtraConstraints} can be linearized.  In the next section, we reformulate these optimization problems in an attempt to speedup computations.

\subsection{Graph reductions}
\label{app:graph}
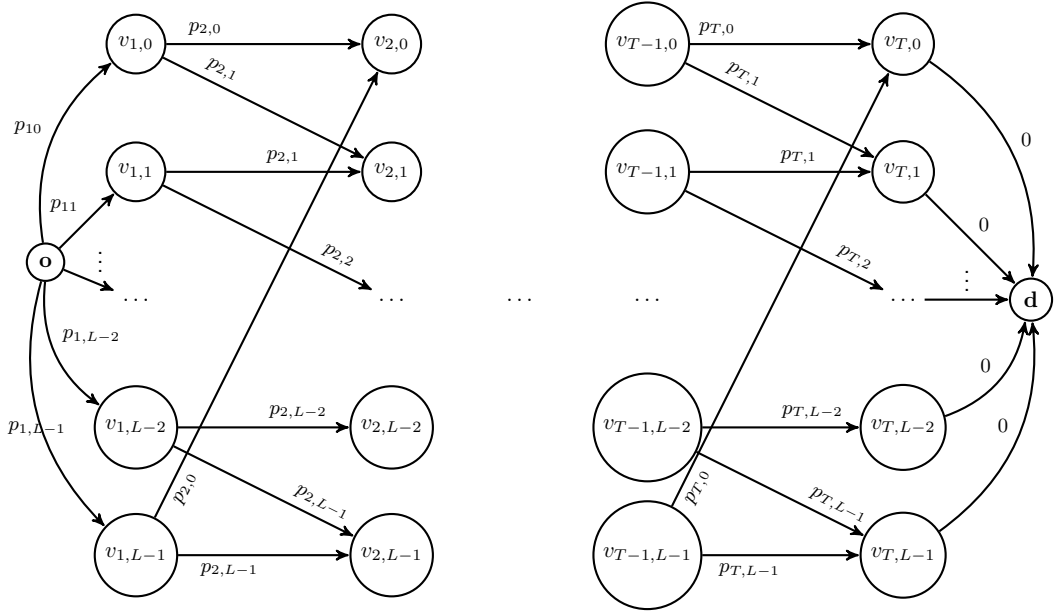
\begin{figure*}
\centering
\begin{tikzpicture}[-,>=stealth',shorten >=0.3pt,auto,node distance=2cm,
  thick,stateNode node/.style={circle,draw},EmptyNode node/.style={draw=none,fill=none}, scale=.85, transform shape]
   \tikzstyle{matched} = [draw,line width=3pt,-]

  \node[stateNode node] (1) {$\mathbf{o} $};
  \node[stateNode node] (2) [above right of=1] {$v_{1,1} $};
    \node[stateNode node] (3) [above  of=2] {$v_{1,0} $};
  \node[EmptyNode node] (4) [below of=2] {$\ldots $};
   \node[stateNode node] (5) [below  of=4] {$v_{1,L-2} $};
      \node[stateNode node] (10) [below  of=5] {$v_{1,L-1} $};
   \node[stateNode node,  node distance=4cm] (6) [right of=3] {$v_{2,0} $};
    \node[stateNode node] (7) [below  of=6] {$v_{2,1} $};
  \node[EmptyNode node] (8) [below of=7] {$\ldots $};
   \node[stateNode node] (9) [below  of=8] {$v_{2,L-2} $};
     \node[stateNode node] (11) [below  of=9] {$v_{2,L-1} $};
      \node[EmptyNode node] (12) [right  of=8] {$\ldots $};
    \node[stateNode node,  node distance=4cm] (13) [right of=6] {$v_{T-1,0} $};
    \node[stateNode node] (14) [below  of=13] {$v_{T-1,1} $};
  \node[EmptyNode node] (15) [below of=14] {$\ldots $};
   \node[stateNode node] (16) [below  of=15]  {$v_{T-1,L-2} $};
     \node[stateNode node] (17) [below  of=16] {$v_{T-1,L-1} $};
     \node[stateNode node,  node distance=4cm] (18) [right of=13] {$v_{T,0} $};
    \node[stateNode node] (19) [below  of=18] {$v_{T,1} $};
  \node[EmptyNode node] (20) [below of=19] {$\ldots $};
   \node[stateNode node] (21) [below  of=20]  {$v_{T,L-2} $};
     \node[stateNode node] (22) [below  of=21] {$v_{T,L-1} $};
     \node[stateNode node] (23) [right  of=20] {$\mathbf{d} $};

  \path[->,every node/.style={font=\sffamily\small}]
   (1) edge [bend left] node  {$p_{10}$} (3)
         edge node  {$p_{11}$} (2)
         edge node  {$\vdots$} (4)
         edge [bend right]  node  {$p_{1,L-2}$} (5)
         edge [bend right]  node[below]  {$p_{1,L-1}$} (10)
    (3) edge node[pos=0.15,sloped]  {$p_{2,1}$} (7)
         edge node[pos=0.2,sloped]  {$p_{2,0}$} (6)
   (2)  edge node [pos=0.6,sloped]  {$p_{2,1}$} (7)
         edge  node[pos=0.7,sloped]  {$p_{2,2}$} (8)
   (5)  edge  node [pos=0.7,sloped] {$p_{2,L-2}$} (9)
           edge  node[pos=0.6,sloped]  {$p_{2,L-1}$} (11)
    (10)  edge node[pos=0.3,sloped,below]  {$p_{2,L-1}$} (11)
           edge  node[pos=0.08,sloped,below]  {$p_{2,0}$} (6)
   (13) edge node[pos=0.15,sloped]  {$p_{T,1}$} (19)
         edge node[pos=0.15,sloped]  {$p_{T,0}$} (18)
   (14)  edge  node [pos=0.6,sloped]  {$p_{T,1}$} (19)
         edge  node[pos=0.7,sloped]  {$p_{T,2}$} (20)
   (16)  edge  node [pos=0.7,sloped] {$p_{T,L-2}$} (21)
           edge  node[pos=0.6,sloped]  {$p_{T,L-1}$} (22)
    (17)  edge node[pos=0.3,sloped,below]  {$p_{T,L-1}$} (22)
           edge  node[pos=0.06,sloped,below]  {$p_{T,0}$} (18)
       (18) edge [bend left] node  {$0$} (23)
        (19)  edge node  {$0$} (23)
       (20)  edge node  {$\vdots$} (23)
       (21)  edge [bend right]  node  {$0$} (23)
        (22) edge [bend right]  node  {$0$} (23);
\end{tikzpicture}
\caption{Graph representing allowed probabilistic state transition.}
\label{fig:LongestPath}
\end{figure*}

We start by mapping problem~\eqref{QuadraticOPT}  in a longest path problem in a directed acyclic graph $G=(V,A)$\footnote{Note that by changing the sign of the distances in the graph, it becomes a shortest path problem.}:
\begin{itemize}
    \item Set of vertices: $V= \lbrace \mathbf{o}\rbrace \cup \lbrace v_{ts}, \textrm{ for } t=1,\ldots,T, \textrm{ for } s \in S \rbrace  \cup\lbrace \mathbf{d} \rbrace$, where $S= \lbrace 0,1, \ldots, L-1\rbrace$.
    \item Set of arcs be $A= \lbrace  (\mathbf{o}, v_{1s}), \textrm{ for } s \in S   \rbrace  \cup \mathcal{A} \cup \lbrace  ( v_{Ts}, \mathbf{d}), \textrm{ for } s \in S   \rbrace$,  where $\mathcal{A}= \lbrace  (v_{ts}, v_{t+1,s}), (v_{ts}, v_{t+1,s+1}),  \textrm{ for } t=1,\ldots,T-1, \textrm{ for } s \in S   \rbrace$. 
    \item Distances $d_a$ for arcs $a \in A$: $(\mathbf{o}, v_{1s})$ and  $( v_{Ts}, \mathbf{d})$ have distance $p_{1s}$ and $0$, respectively, and  $(v_{t-1,s}, v_{ts'})$  has distance $p_{ts'}$.
\end{itemize}
In Figure~\ref{fig:LongestPath}, it is illustrated the defined graph, note that the arcs enforce Constraints~\eqref{const2}. In this way, a solution of Problem~\eqref{QuadraticOPT} is equivalent to the computation of the longest path between the origin $\mathbf{o}$ and the destiny $\mathbf{d}$ in this graph. Since the graph is a weighted directed acyclic graph, one can multiply by -1 each distance, reducing the problem to a shortest path (with negative weights) which is well-known to be solvable in polynomial time, e.g., using Bellman-Ford algorithm~\cite{BELLMAN1958,Ford:1962}. 

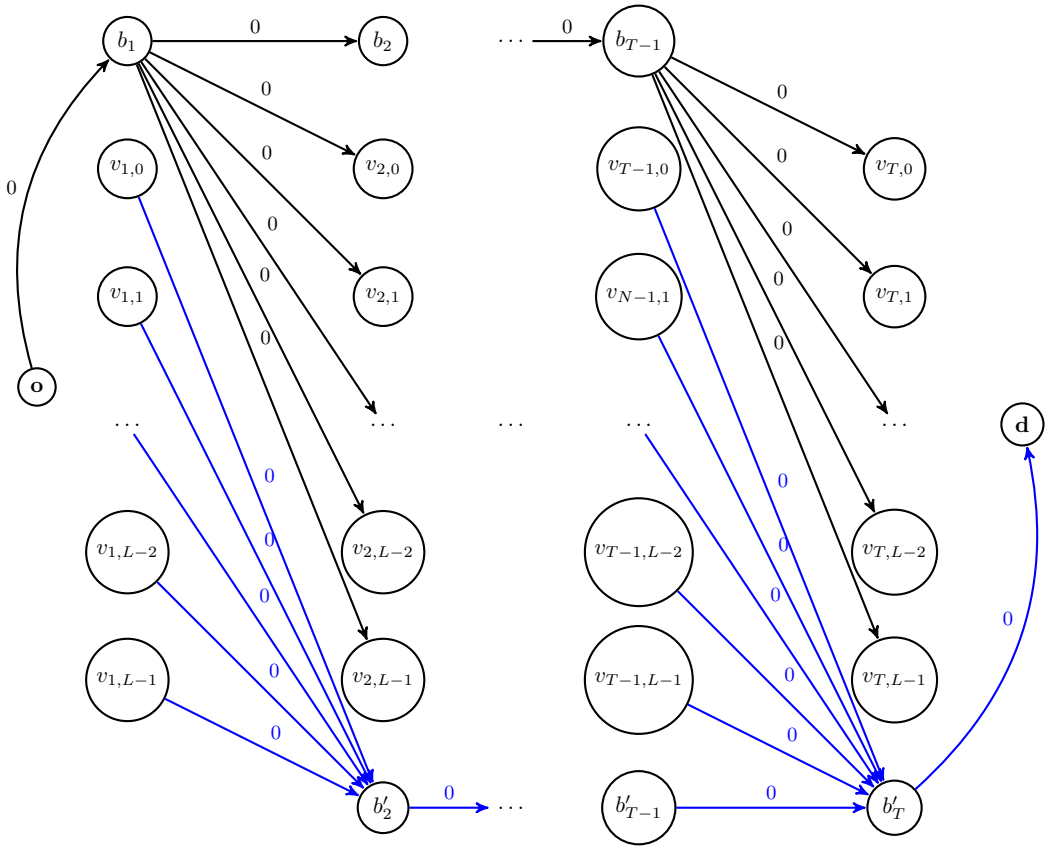
\begin{figure*}
\centering
\begin{tikzpicture}[-,>=stealth',shorten >=0.3pt,auto,node distance=2cm,
  thick,stateNode node/.style={circle,draw},EmptyNode node/.style={draw=none,fill=none}, scale=.85, transform shape]
   \tikzstyle{matched} = [draw,line width=3pt,-]

  \node[stateNode node] (1) {$\mathbf{o} $};
  \node[stateNode node] (2) [above right of=1] {$v_{1,1} $};
    \node[stateNode node] (3) [above  of=2] {$v_{1,0} $};
    \node[stateNode node] (24) [above  of=3] {$b_1$};
  \node[EmptyNode node] (4) [below of=2] {$\ldots $};
   \node[stateNode node] (5) [below  of=4] {$v_{1,L-2} $};
      \node[stateNode node] (10) [below  of=5] {$v_{1,L-1} $};
   \node[stateNode node,  node distance=4cm] (6) [right of=3] {$v_{2,0} $};
   \node[stateNode node] (25) [above of=6] {$b_2$};
    \node[stateNode node] (7) [below  of=6] {$v_{2,1} $};
  \node[EmptyNode node] (8) [below of=7] {$\ldots $};
   \node[stateNode node] (9) [below  of=8] {$v_{2,L-2} $};
     \node[stateNode node] (11) [below  of=9] {$v_{2,L-1} $};
     \node[stateNode node] (29) [below of=11] {$b'_2$};
      \node[EmptyNode node] (12) [right  of=8] {$\ldots $};
    \node[stateNode node,  node distance=4cm] (13) [right of=6] {$v_{T-1,0} $};
    \node[stateNode node] (26) [above of=13] {$b_{T-1}$};
    \node[stateNode node] (14) [below  of=13] {$v_{N-1,1} $};
  \node[EmptyNode node] (15) [below of=14] {$\ldots $};
   \node[stateNode node] (16) [below  of=15]  {$v_{T-1,L-2} $};
     \node[stateNode node] (17) [below  of=16] {$v_{T-1,L-1} $};
     \node[stateNode node] (30) [below of=17] {$b'_{T-1}$};
     \node[stateNode node,  node distance=4cm] (18) [right of=13] {$v_{T,0} $};
    \node[stateNode node] (19) [below  of=18] {$v_{T,1} $};
  \node[EmptyNode node] (20) [below of=19] {$\ldots $};
   \node[stateNode node] (21) [below  of=20]  {$v_{T,L-2} $};
     \node[stateNode node] (22) [below  of=21] {$v_{T,L-1} $};
      \node[stateNode node] (31) [below of=22] {$b'_{T}$};
     \node[stateNode node] (23) [right  of=20] {$\mathbf{d} $};
     \node[EmptyNode node] (32) [left of=26] {$\ldots $};
      \node[EmptyNode node] (33) [right of=29] {$\ldots $};

  \path[->,every node/.style={font=\sffamily\small}]
   (1)  edge [bend left]  node  {$0$} (24)
   (24)  edge  node  {$0$} (25)
         edge  node  {$0$} (6)
         edge  node  {$0$} (7)
         edge  node  {$0$} (8)
         edge  node  {$0$} (9)
         edge  node  {$0$} (11)
   (32)  edge  node  {$0$} (26)
   (26)  edge  node  {$0$} (18)
         edge  node  {$0$} (19)
         edge  node  {$0$} (20)
         edge  node  {$0$} (21)
         edge  node  {$0$} (22);
  \path[->,every node/.style={font=\sffamily\small},color=blue]
   (3)  edge  node  {$0$} (29)
   (2)  edge  node  {$0$} (29)
   (4) edge  node  {$0$} (29)
   (5) edge  node  {$0$} (29)
   (10) edge  node  {$0$} (29)
   (29) edge  node  {$0$} (33)
   (30) edge  node  {$0$} (31)
   (14) edge  node  {$0$} (31)
   (13) edge  node  {$0$} (31)
   (15) edge  node  {$0$} (31)
   (16) edge  node  {$0$} (31)
   (17) edge  node  {$0$} (31)
   (31) edge [bend right]  node  {$0$} (23);
\end{tikzpicture}
\caption{Graph adaptation for Problem~\eqref{ExtraConstraints} with only the new arcs $A'$.}
\label{fig:LongestPathExtra}
\end{figure*}

Replicating the same reasoning, we can adapt Problem~\eqref{ExtraConstraints} to  a restricted shortest path problem by adding to $G=(V,A)$:
\begin{itemize}
    \item The set of vertices $\lbrace b_t, \textrm{ for } t=1,\ldots,T-1 \rbrace \cup \lbrace b'_t, \textrm{ for } t=2,\ldots,T \rbrace$. 
    \item The set of arcs $A'=\lbrace (\mathbf{o},b_1) \rbrace \cup \lbrace (b'_T,\mathbf{d}) \rbrace \cup \lbrace (b_t,v_{t+1,s}), \textrm{ for } t=1,\ldots,T-1,  \textrm{ for } s \in S \rbrace \cup \lbrace (b_t,b_{t+1}), \textrm{ for } t=1,\ldots,T-2 \rbrace \cup \lbrace (v_{ts},b'_{t+1}), \textrm{ for } t=1,\ldots,T-1,  \textrm{ for } s \in S \rbrace \cup \lbrace (b'_t,b'_{t+1}), \textrm{ for } t=2,\ldots,T-1 \rbrace$.
    \item Distance $d_a=0$ for all $a \in A'$ (new arcs).
\end{itemize}
In Figure~\ref{fig:LongestPathExtra}, it is illustrated part of the new defined graph (for sake of simplicity, we only draw the new arcs $A'$). 
By construction, once a path passes from a vertex $b_t$ to a $v_{t+1,s}$, it can not go back to any vertex $b$. Analogously, once a path goes from a vertex $v_{ts}$ to  $b'_{t+1}$, it cannot go back to any vertex $v$.

The constrained shortest path problem becomes:
\begin{subequations}
\begin{alignat}{5}
&\min_y  & \sum_{a \in A} -d_a y_{a}\label{Obj1_short}\\
& s.t. & \sum_{a \in \delta^+(\mathbf{o})} y_a =1  \label{const_o}\\
& &  \sum_{a \in \delta^-(\mathbf{d})} y_a =1  \label{const_d}\\
& & \sum_{a \in \delta^-(\mathbf{v})} y_a = \sum_{a \in \delta^+(\mathbf{v})} y_a  & \qquad \forall v \in V \setminus \lbrace \mathbf{o}, \mathbf{d} \rbrace \label{const_flow}\\
&      & \sum_{a \in A}  y_{a} = J F&  \label{const_5f}\\
& & y_{a} \in \lbrace 0,1 \rbrace & \qquad \forall a \in A,\label{const_binary}
\end{alignat}
\label{Shortest_path_ExtraConstraints}
\end{subequations}
where $\delta^+(\mathbf{v})$ corresponds to the outgoing arcs of v,  $\delta^-(\mathbf{v})$ corresponds to the incoming arcs of v, $y_a$ represents the arcs selected in the shortest path. Constraint~\eqref{const_o} and Constraint~\eqref{const_d} enforce that the path starts in $\mathbf{o}$ and ends in  $\mathbf{d}$, respectively. Constraints~\eqref{const_flow} are the standard flow conservation constraints. Constraint~\eqref{const_5f} enforces that at least $JF_s$ arcs of $A$ (initial graph) are used. While efficient algorithms exist for determining a shortest path in a graph, constrained shortest path problems are NP-hard~\cite{Garey:1979}. Nevertheless, several solution strategies exist to solve them in practice: parallel methods, dynamic programming and Lagrangian relaxation, see for example~\cite{LOZANO2013,ZHU2012,Carlyle2008}. 

The integer programming problems of the previous sections reflect the flexibility for these models to embed physiological constraints and clinical knowledge. Furthermore, these combinatorial optimization models also offer flexibility in terms of the user optimization criteria. For instance, the objective function could be replaced by the entropy function.

\subsection{Comparison of formulations}

In what follows, we briefly discuss the complexity associated with solving the optimization problems of the previous sections. Table~\eqref{table:size_problems} summarizes the comparison between the sizes of our formulations.
Although Problems~\eqref{QuadraticOPT} and~\eqref{ExtraConstraints} can be  linearized, they still have binary variables which might make their direct resolution prohibitive for long recordings. For this reason, in Section~\ref{app:graph}, we provided alternative formulations which in practice are solved more efficiently. We reduced Problem~\eqref{QuadraticOPT} to a shortest path problem in an acyclic graph which is known to be solvable in polynomial time. This result implies that Problem~\eqref{ExtraConstraints} can be solved by determining for each possible consecutive $J$ seconds the shortest path, i.e., solving the shortest path problem in $T-JF+1$ acyclic graphs. Alternatively, Problem~\eqref{ExtraConstraints} is equivalent to solving a \textit{constrained} shortest path problem, Problem~\eqref{Shortest_path_ExtraConstraints}. Although from Table~\eqref{table:size_problems}, we conclude that Problem~\eqref{Shortest_path_ExtraConstraints} is the one with more binary variables, in our computational experiments  with Gurobi\footnote{Gurobi: www.gurobi.com}, it was observed that solving it required few seconds, while Problem~\eqref{ExtraConstraints}  demanded several minutes. This is explained by the much tighter relaxation that Problem~\eqref{Shortest_path_ExtraConstraints} provides in comparison with Problem~\eqref{ExtraConstraints}. For this reason, in all the experiments reported in this paper, we used the graph reformulations presented in Section~\ref{app:graph}.

\begin{table} \hspace{-0.4cm}
\begin{tabular}{m{\mylength}|m{\mylength}|m{\mylength}|m{\mylength}|} \footnotesize
  &  Number of  & Number of & Number of \\
  &  Var. & Binary Var. & Constraints \\ \hline
Prob.~\eqref{QuadraticOPT}  & TL  & TL& 2T-1\\
Lin. of Prob.~\eqref{QuadraticOPT}     &  TL+2L(T-1)& TL &2T-1+6L(T-1)\\
Prob.~\eqref{ExtraConstraints} & TL+T&   TL+T & 2T+1 \\
Lin. of Prob.~\eqref{ExtraConstraints} &TL+T+(T-1)(2+2L)+1 & TL+T &  2T+1+3(2T-1)+6L(T-1)\\
Prob.~\eqref{Shortest_path_ExtraConstraints} & 4TL+2(L+T-1) & 4TL+2(L+T-1) & 2(T-1)+LT+3\\ \hline
\end{tabular}
\caption{Size of the combinatorial optimization problems. Lin. and Var. stand for linearization and variables, respectively.}
\label{table:size_problems}
\end{table}

\section{Case Study: Heart Sound Segmentation}\label{PCG_CASE_STUDY}



\subsection{PCG morphology and characteristics}
In each heart beat, two distinct waves are produced by the heart. When the atrioventricular valves close, the resulting wave, known as the first heart sound (S1) is low in pitch and relatively long-lasting \cite{hall2010guyton}. The next stage, corresponds to the systolic period where large amounts of blood are injected from the ventricles to the pulmonary and aortic arteries. At the end of this period, some expected blood flows back to the ventricles, forcing the aortic and pulmonary valves to close, as a result, a rapid snap sound called the second heart sound (S2) is generated. Finally, in the last stage the ventricles relax, and are filled once again with blood, a period known as diastolic period \cite{hall2010guyton}. The S1 and S2 sounds are recorded by a digital stethoscope and the corresponding audio signal is known as the PCG signal, see Figure \ref{ECG_PCG_signals}.

\subsection{Why is auscultation important?} Cardiovascular diseases (CVDs) are the leading cause of death in developed and developing countries and one of the major causes of hospitalization. By 2030, almost 23.6 million people will die from CVDs, according to the world health organization  \cite{Wealth:2011}. One major solution, goes by an effective screening of the population, not only to identify risk groups but also to forward immediately those who need emergent care. In this sense, heart sound auscultation represents a key exam, due to its simplicity and low cost, that can be used as a first line of screening for several heart diseases, including arrhythmia, valve diseases, heart failure, etc.  One of such steps, concerns the detection of the fundamental heart sounds (S1) and (S2) and also the detection of systolic and diastolic periods. In this section, our proposed solution is going to decode the existence of such a waves, in a wide variety of pathogenic cases.

\begin{figure}[t]
  \centering
  \includegraphics[width=0.50\textwidth]{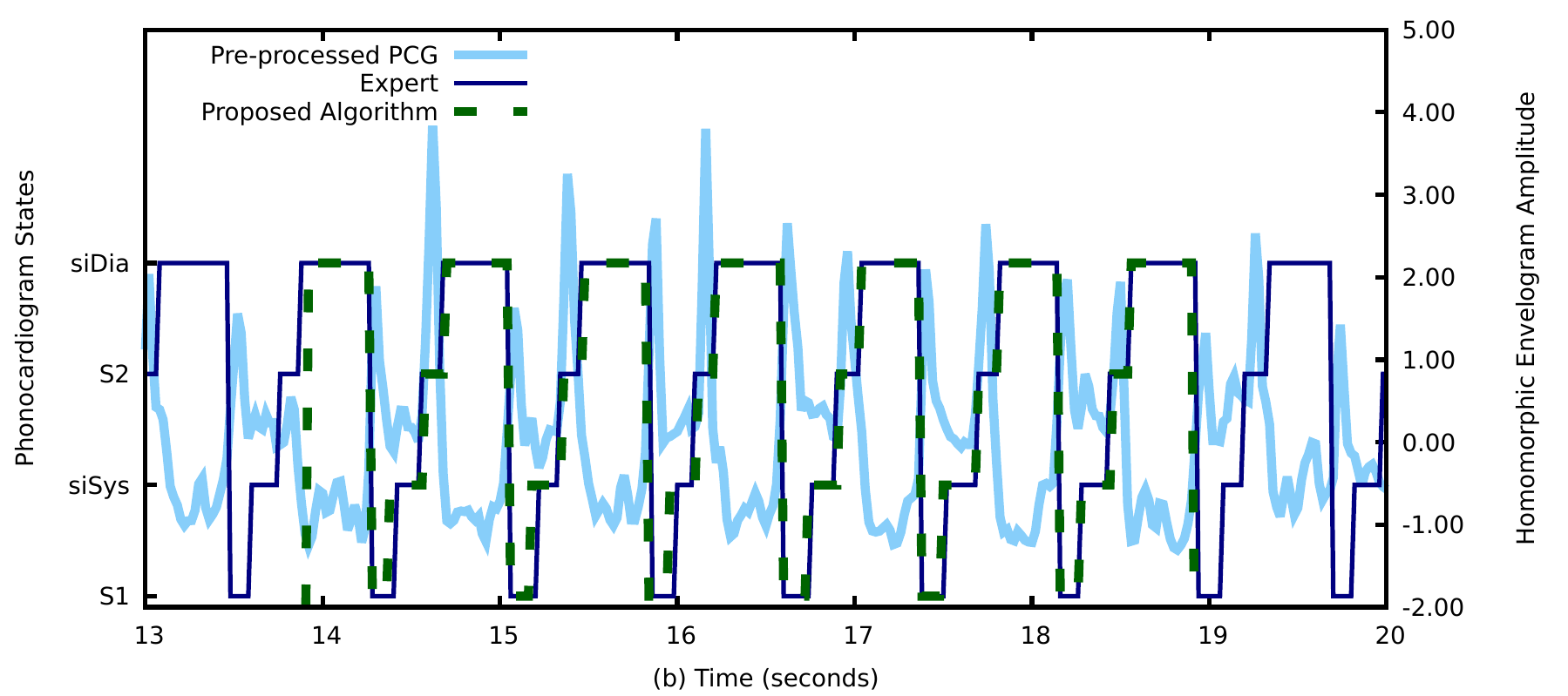}
  \caption{An example of the proposed segmentation algorithm. A pre-processed heart sound signal (light blue line) with the corresponding manual annotations provided by Physionet/Cinc Challenge (dark blue line) are displayed. The algorithm displays (green dash line) of what it considers to be the \enquote{best} five seconds of the recording, according to the criteria defined in equation \eqref{Obj1}. }
  \label{fig:partially_segmented_signal}
\end{figure}

\subsection{Materials}\label{Materials_Chapter}
In this work, the database from the 2016 PhysioNet/ Computing in Cardiology Challenge~\cite{Physionet2} is used. The database provides a large collection of heart sound recordings, divided into eight different training sets. The heart sound databases included in the Physionet dataset were collected independently by seven different research teams. Therefore, hardware, recording spots, data quality and target population are substantially different.
The Physionet/Cinc Challenge provided 3153 heart sounds. From this, 363 heart sounds were discarded by the following reasons: 87 records from Folder E are not heart sounds; 276 records do not have annotations or they are not properly annotated, therefore not considered in this study.  \\
Although signals were collected at different sampling frequencies (800Hz, 1000Hz, 2000 Hz, 3000Hz, 4000 Hz, 8000 Hz or 22050 Hz) no details concerning aliasing and imaging effects are provided.

\subsection{Training and Testing LSTM models}
Aiming to get statistically significant results a large dataset is created. To do so, the different constituent datasets of the 2016 Physionet/Computers in Cardiology Challenge (discussed in Section \ref{Materials_Chapter}) are merged.
During the process, $70\%$ of healthy and unhealthy patients (and their corresponding audio records) are randomly used for training  and the remaining ones are used for testing. 
In order to avoid over-fitting, recordings from the same subject are only used for training or for testing purposes. Furthermore, $10\%$ of the training data is randomly used for validation purposes. In order to avoid biasing, recordings of the same subject are only used to train the model or to measure the fitting quality of the model, respectively.
After the data has been split into train, validation and test sets, the following steps are applied in each corresponding set:
\begin{itemize}
\item Following previous literature~\cite{Gupta:2007,Gill:2005}, the PCG signal is first normalized into the range $[0,1]$, through a min-max normalization method \cite{Bandyopadhyay_2014}, as it is shown in equation (\ref{tilda_equation}),
\begin{equation}
\tilde{x}_{t} = \frac{x_{t}-\mathrm{min}(X)}{\mathrm{max}(X)-\mathrm{min}(X)},
\label{tilda_equation} 
\end{equation}where $\tilde{x}_{t}$ is used to denote normalized samples. 
\item Then, the signal is filtered using Butterworth lowpass and highpass filters of order $4$, with cutoff frequencies at $400$ Hz and $25$ Hz, respectively.
\item Afterwards, the homomorphic envelogram,  the Hilbert envelogram, the wavelet envelope and the power spectral density (PSD) envelope are extracted from the filtered signal, as in \cite{Springer:2016}. 
\end{itemize}
Moreover, similarly to \cite{Springer:2016}, such envelopes are further downsampled to 50 Hz, in order to reduce the computational complexity of the segmentation method.\\
Before starting the training phase, $L$ is going to be set to four (S1, Systolic, S2 and Diastolic) states and $N$ is going to be set to four (homomorphic envelogram, Hilbert envelogram, wavelet envelope, power spectral density envelope) features and $M$ is fixed to 32 in the current case.
The matrices $\pmb{W_{x \cdot}}$, $\pmb{W_{h \cdot}}$ $\pmb{W_{out}}$ and the bias vectors $b_{i}$, $b_{f}$, $b_{o}$, $b_{j}$ are initialized using a uniform random distribution over the domain $[ -0.05 , 0.05 ]$. The $h$  and $c$ vectors are zero initialized, i.e,  all components are equal to zero. The NN is trained during 25 epochs, aiming to maximize the binary cross entropy. To do so, the Adam optimizer proposed by \cite{Kingma:2014} is used, where the learning rate and $\epsilon$ are set to $0.01$ and $10^{-6}$ respectively, during the entire learning phase. At the beginning of each epoch, signals are sorted randomly from a uniform random distribution over the domain $[1,N_s]$, where $N_s$ is the size of the train dataset. Each signal is analysed individually and sequentially (patches of size one). Furthermore, the NN weights are saved at the end of each epoch. At the very end of the training phase, only the NN weights that achieves the lowest loss value (in our case mean square error) in the cross-validation dataset is saved to be further evaluated during the testing phase. Finally, in order to extract statistically significant results, the aforementioned procedure is repeated 10 times. Note, that the generated training, validation and testing sets at each run are statistically different from each other, i.e, the ratio of healthy and unhealthy records is different at each trial.


\subsection{Performance Metrics}\label{PerformanceMetrics}
The simplest performance metric used in this work is sample accuracy ($A$), which represents the fraction of samples in the output state sequence that are correctly allocated to the corresponding state in the ground truth state sequence. Other two metrics concern the detection of the fundamental heart sounds $S1$ and $S2$: specificity (Spec) and sensitivity (Sens). A true positive exists when the center of an S1 (S2) sound in the estimated state sequence is closer than 60 ms from the center of the corresponding S1 (S2) sound in the ground truth state sequence. All the others S1 and S2 sounds in the estimated state sequence are considered to be false positives. On the other hand, a true negative happens when the center of an Systolic (Diastolic) state in the estimated state sequence is closer than 60 ms from the center of the corresponding Systolic (Diastolic) state in the ground truth state sequence. Finally, the average performance is computed over the entire test set for each of the ten trials, and reported in Section \ref{Results}.

\subsection{Results}\label{Results}

\begin{figure}
\centering
\begin{tikzpicture}
\begin{axis}[
ymax=1,
ymin=0.9,
xmax=3,
xmin = 0,
boxplot/draw direction=y,
cycle list={{ultra thin}, {ultra thick},{thick}},
legend cell align={right}, 
        legend pos= south east,
        legend image post style={sharp plot},
boxplot={
       draw position={
                1/4 + floor(\plotnumofactualtype/3)
                  + 1/4*fpumod(\plotnumofactualtype,3)
            },
            box extend=0.1,},
xtick={0,1,2,3},
x tick label as interval,
xticklabels={ {$A$},{Sens},{Spec}},
		label style={font=\small},
        x tick label style={font=\small},
]      
]

\addplot+[mark = x, mark options = {mark color=grey},boxplot]
table[y index=0] {accuracy_naive.txt};

\addplot+[mark = x, mark options = {mark color=grey},boxplot]
table[y index=0] {accuracy_opt.txt};

\addplot+[mark = x, mark options = {mark color=grey},boxplot]
table[y index=0] {accuracy_naive_windows.txt};

\addplot+[mark = x, mark options = {mark color=grey},boxplot]
table[y index=0] {sensitivity_naive.txt};

\addplot+[mark = x, mark options = {mark color=grey},boxplot]
table[y index=0] {sensitivity_opt.txt};

\addplot+[mark = x, mark options = {mark color=grey},boxplot]
table[y index=0] {sensitivity_naive_windows.txt};

\addplot+[mark = x, mark options = {mark color=grey},boxplot]
table[y index=0]{specificity_naive.txt};

\addplot+[mark = x, mark options = {mark color=grey},boxplot]
table[y index=0]{specificity_opt.txt};

\addplot+[mark = x, mark options = {mark color=grey},boxplot]
table[y index=0]{specificity_naive_windows.txt};

\draw[dotted] (axis cs: 1,\pgfkeysvalueof{/pgfplots/ymin}) -- 
                      (axis cs: 1,\pgfkeysvalueof{/pgfplots/ymax});
                      
   \draw[dotted] (axis cs: 2,\pgfkeysvalueof{/pgfplots/ymin}) -- 
                      (axis cs: 2,\pgfkeysvalueof{/pgfplots/ymax});
   
   \draw[dotted] (axis cs: 3,\pgfkeysvalueof{/pgfplots/ymin}) -- 
                      (axis cs: 3,\pgfkeysvalueof{/pgfplots/ymax}); 
                                           
\end{axis}
\end{tikzpicture}
\caption{Heart sound segmentation results on the Physionet dataset. An algorithm, which follows the standard method to assign a state to a sample (thin solid lines),
see Section \ref{Standard_Approaches} for more details. An algorithm which assigns a state to a sample by finding the optimal solution of the problem \eqref{ExtraConstraints} (thick solid line), for more details see Section \ref{G_otimization}. An algorithm which assigns a state to a sample, by first finding the optimal window, which satisfies the problem \eqref{G_otimization}, and then outputs a state sequence inside of the optimal window, by following the traditional methods presented in Section \ref{Standard_Approaches}, the results are displayed in solid line. } \label{fig:M7}
\end{figure}
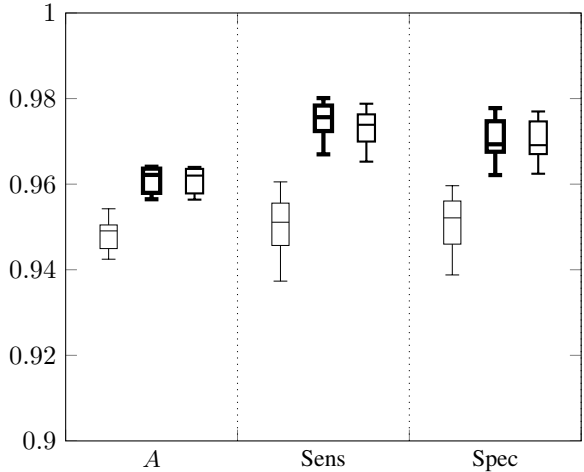
In this section, the impact of the proposed methods are measured and reported in Figure \ref{fig:M7}. The performance of an algorithm, which follows the standard approach, i.e. a state is assigned at a particular time instance according to the equation (\ref{argmax_k}), is used as a baseline for comparisons. 
Furthermore, in order to make fair comparisons, 
random windows of the same size as the one proposed by the algorithm described in Section \ref{G_otimization} are created and allocated, as a result, only the states inside of the interest window are further evaluated, the results are displayed in Figure \ref{fig:M7} in thin solid line. \\
In our first experiment, an algorithm attempts to accomplish two tasks simultaneously: it searches for the most likely state sequence of events according to physiological constraints in the cardiac cycle and it also searches for the optimal region of interest for a further post processing stage, the results are displayed in Figure \ref{fig:M7} in thick line.
The results show a significant boost in performances, on average $A$, Sens and Spec increases $1,3\%$, $2,4\%$ and $1,9\%$ respectively. \\ It is not very likely that all audio segments have the same quality, e.g it is expected that the first and the last segments of a recording have a low signal-to-noise ratio due to rubbing noise, generated by the stethoscope in contact with the human skin. It is also very common a person to cough, talk or move during an auscultation, and as a result,  noisy audio segments are likely to happen anywhere in the recording. The algorithm surpasses, these troubles by automatically selecting the optimal region for a further post-processing stage, according to a criteria defined by the user. 
In our second experiment, the effect of the optimal window is measure independently if cardiac transition constraints are satisfied or not. To do so, the interest window is obtained by solving the non-linear problem  in \eqref{ExtraConstraints}. But instead of returning the corresponding state output variables, it returns the sequence generated by applying an argmax function, column-by-column to the state output probability distribution, as it is explained in Section \ref{Standard_Approaches}. The results are displayed in Figure \ref{fig:M7} in solid lines. As it is possible to observe, the results are not different when compared to our first experiment. This might be due to the fact, that indeed inside of this interest window, both algorithms are very confident in assigning a state to a sample, and perhaps, it is inside of this window, where state transitions constraints are easily obeyed by both approaches. 

\section{Case Study: Electrocardiogram Segmentation}\label{ECG_CASE_STUDY}

\subsection{ECG morphology and characteristics}
\begin{figure}
\centering\includegraphics[scale=0.5]{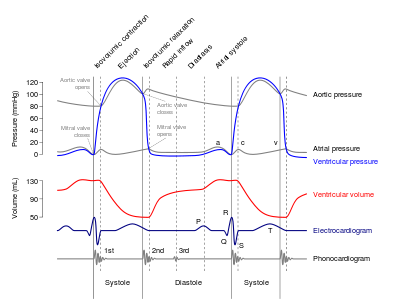}\caption[]{ The Wiggers diagram, including ECG and PCG signals at the bottom of the figure (adapted from \cite{Hall:2011}).}
\label{ECG_PCG_signals}
\end{figure}

When a cardiac impulse spreads through the heart, electromagnetic waves also spread from the heart into the adjacent tissues surrounding the heart. These are detected and recorded by electrodes placed on opposite sides of the heart. This recording is known as an ECG signal. 
A normal ECG (see Figure \ref{ECG_PCG_signals}) is composed by a P, QRS complex, and T waves. The P wave is caused by electrical potentials generated when the atrium depolarize. The QRS complex is caused by electrical potentials generated when the ventricles depolarize. The T wave is caused by electrical potentials generated as the ventricles recover from the state of depolarization \cite{Hall:2011}. Between these electromagnetic waves, equipotential lines exist.

\subsection{Why the ECG analyzes is important?}
The ECG is another important heart signal, and it provides information concerning the heart's rhythm and electrical activity \cite{hall2010guyton}. An ECG is often used alongside with other  exams to diagnose arrhythmias, coronary heart disease, cardiomyopathy, etc. The ECG analysis, is one of the most common procedures, and several systems have been developed aiming to provide ECG diagnosis. To do so, one of such steps, concerns the detection of electro-magnetic waves, such as QRS complex, P and T-wave, etc. In this section, our proposed solution is going to decode the existence of such waves in a wide variety of cases. 


\subsection{Materials}
In this work, the dataset QT is used to measure the impact of our propose solutions when segmenting ECG signals.  In this database, a wide variety of QRS and ST-T morphologies are available. The dataset compresses a total of 105 fifteen-minute excerpts of two ECG channels, recorded at 250Hz.  Within each record, between 30 to 100 heart beats were manually annotated by cardiologists, who identified the beginning, the peak and the ending of the P-wave, the beginning and the ending of the QRS complex, the peak and the ending of the T-wave, and (if present) the peak and the ending of the U-wave, although this last wave was not considered in this study.  In all, 3622 beats have been annotated by cardiologists. These annotations have been carefully audited to eliminate gross errors, although the precise placement of each annotation was left to the judgment of expert annotators.

\subsection{Training and Testing LSTM models}
The QT dataset is relatively small, only 105 independent records of 15 minutes long are at disposal. 
As a result, ECG signals are split into continuous segments of 30 seconds, thus generating 3150 dependent ECG signals. Furthermore, 70\% of the ECG recordings are randomly used for training and the remaining ones are used for testing. In order to avoid over-fitting, recordings from the same subject are only used for training or for testing purposes.  Furthermore, $10\%$ of the training data is randomly used for validation purposes. To avoid biasing, recordings of the same subject are only used to train the model or to measure its fitting quality of in the cross-validation dataset respectively. 
After the dataset has been split into train, validation and test sets, raw ECG signals are standardized, so the data has zero mean and unitary variance. Furthermore, seven distinct features are extracted:
\begin{itemize}
    \item The QRS envelogram proposed by \cite{Pan:1985}. In this transformation, a cascade of filters is applied in order to attenuate noise and to enhance QRS complex waves.
    \item Wavelet based envelograms. In this paper, a one-dimensional discrete stationary wavelet transform is applied to the standardized signal, using as a prototype wavelet the quadratic spline function, proposed by \cite{Mallat:1992}. Following \cite{Martinez:2004}, the first six approximation coefficients are computed and saved. Our choices are based on the following arguments: decomposition's up to $2^4$, the energy of the QRS complex wave dominates the energies from the P and T wave respectively. On the other hand decomposition's above $2^4$, P and T waves are expected to have a significant higher energy contribution than for example QRS complex or U waves \cite{Martinez:2004}.
\end{itemize} These envelograms are further downsampled to 50 Hz and once more standardized.
Before starting the training phase, $L$ is going to be equal to six (P-wave, interval between a P and QRS-wave, QRS-wave, interval between QRS and a T-wave, T-wave and finally the interval between a T and a P-wave) states and $N$ is going to be equal to seven (one QRS envelogram and six wavelet envelograms) features and $M$ is fixed to 32 in the current case. the matrices $\pmb{W_{x \cdot}}$, $\pmb{W_{h \cdot}}$ $\pmb{W_{out}}$ and the bias vectors $b_{i}$, $b_{f}$, $b_{o}$, $b_{j}$ are initialized using a uniform random distribution over the domain $[ -0.05 , 0.05 ]$. The $h$  and $c$ vectors are zero initialized. The NN is trained during 25 epochs, aiming to maximize the binary cross entropy function. To do so, the Adam optimizer proposed by \cite{Kingma:2014} is used, where the learning rate and $\epsilon$ are set to $0.01$ and $10^{-6}$ respectively, during the entire learning phase. At the beginning of each epoch, signals are sorted randomly from a uniform random distribution over the domain $[1,N_s]$. Each signal is analysed individually and sequentially (patches of size one). Furthermore, the NN weights are saved at the end of each epoch. \\
After the model has been trained for 25 epochs, only the NN weights whose loss value computed in the cross-validation dataset is the lowest (in our case mean square error), is saved to be further used during the testing phase. Finally, in order to extract statistically significant results, the aforementioned procedure is repeated 10 times. Note, that the generated train, validation and test sets at each run are statistically different from each other, i.e, the ratio of healthy and unhealthy records is different at each trial.



\subsection{Performance Metrics}
In the case of ECG signals, the same aforementioned metrics are used ($A$, Spec and Sens). A true positive is considered when the center of an QRS complex (P, T) wave in the estimated state sequence is closer than 60 ms from the center of the corresponding QRS complex (P, T) wave in the ground truth state sequence. All other QRS complex, P and T waves in the estimated state sequence are considered to be false positives. On the other hand, a true negative happens for example, when the center of P-QRS (the interval between a P and a QRS complex wave) state in the estimated state sequence is closer than 60 ms from the center of the corresponding P-QRS state in the ground truth state sequence. All performance metrics are computed for each recording in the test set and  averaged over the entire set.  Finally, the average performance is computed over the entire test set for each of the ten trials, and reported in Section  \ref{Results_ECG}.

\subsection{Results} \label{Results_ECG}

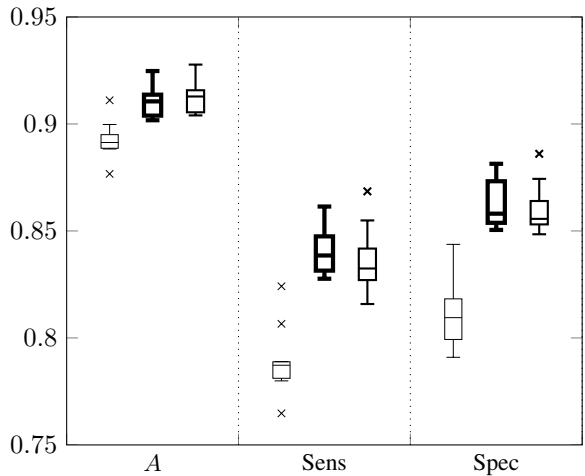
\begin{figure}
\centering
\begin{tikzpicture}
\begin{axis}[
ymax=0.95,
ymin=0.75,
xmax=3,
xmin = 0,
boxplot/draw direction=y,
cycle list={{ultra thin}, {ultra thick},{thick}},
legend cell align={right}, 
        legend pos= south east,
        legend image post style={sharp plot},
boxplot={
       draw position={
                1/4 + floor(\plotnumofactualtype/3)
                  + 1/4*fpumod(\plotnumofactualtype,3)
            },
            box extend=0.1,},
xtick={0,1,2,3},
x tick label as interval,
xticklabels={ {$A$},{Sens},{Spec}},
		label style={font=\small},
        x tick label style={font=\small},
]      
]

\addplot+[mark = x, mark options = {mark color=grey},boxplot]
table[y index=0] {accuracy_naive_ecg.txt};

\addplot+[mark = x, mark options = {mark color=grey},boxplot]
table[y index=0] {accuracy_opt_ecg.txt};

\addplot+[mark = x, mark options = {mark color=grey},boxplot]
table[y index=0] {accuracy_naive_windows_ecg.txt};

\addplot+[mark = x, mark options = {mark color=grey},boxplot]
table[y index=0] {sensitivity_naive_ecg.txt};

\addplot+[mark = x, mark options = {mark color=grey},boxplot]
table[y index=0] {sensitivity_opt_ecg.txt};

\addplot+[mark = x, mark options = {mark color=grey},boxplot]
table[y index=0] {sensitivity_naive_windows_ecg.txt};

\addplot+[mark = x, mark options = {mark color=grey},boxplot]
table[y index=0]{specificity_naive_ecg.txt};

\addplot+[mark = x, mark options = {mark color=grey},boxplot]
table[y index=0]{specificity_opt_ecg.txt};

\addplot+[mark = x, mark options = {mark color=grey},boxplot]
table[y index=0]{specificity_naive_windows_ecg.txt};

\draw[dotted] (axis cs: 1,\pgfkeysvalueof{/pgfplots/ymin}) -- 
                      (axis cs: 1,\pgfkeysvalueof{/pgfplots/ymax});
                      
   \draw[dotted] (axis cs: 2,\pgfkeysvalueof{/pgfplots/ymin}) -- 
                      (axis cs: 2,\pgfkeysvalueof{/pgfplots/ymax});
   
   \draw[dotted] (axis cs: 3,\pgfkeysvalueof{/pgfplots/ymin}) -- 
                      (axis cs: 3,\pgfkeysvalueof{/pgfplots/ymax}); 
                                           
\end{axis}
\end{tikzpicture}
\caption{Electrocardiogram segmentation results, obtained using the QT database \cite{Laguna:1997}. An algorithm, which follows the standard method to assign a state to each sample (thin solid line), see Section \ref{Standard_Approaches} for more details. An algorithm which assigns a state to a sample by finding the optimal solution of the problem \eqref{Shortest_path_ExtraConstraints} (thick solid line), for more details see Section \ref{G_otimization}.  An algorithm which assigns a state to a sample, by first finding the optimal window, which satisfies the problem \eqref{Shortest_path_ExtraConstraints}, and then outputs a state sequence inside of the optimal window, by following the traditional methods presented in Section \ref{Standard_Approaches}, the results are displayed in solid lines. } \label{fig:MC7}
\end{figure}


In this section, the impact of the proposed approaches in segmenting ECGs signals are measured and reported in Figure \ref{fig:MC7}.  The same aforementioned procedure was adopted, i.e, an algorithm which follows the standard approach is used as baseline for comparison reasons. Furthermore, its performance is measure on random windows of 5 seconds long for comparison reasons. In our first experiment, an algorithm attempts to accomplish two tasks simultaneously: it searches for the most likely state sequence of events according to physiological constraints in the cardiac cycle and it also searches for the optimal region of interest for a further post processing stage, the results are displayed in Figure \ref{fig:MC7} in thick lines. As expected, the proposed solution achieved higher $A$, Sens and Spec performances, when compared to the standard approaches presented in Section \ref{Standard_Approaches}. The usage of an optimal window, guarantees on average a boost of $1,8\%$, $4,9\%$ and $4,9\%$ in $A$, Sens and Spec, respectively. As it already happened previously, noisy samples are likely to happen at any moment in ECG signals: electrode contact noise, power-line interference, muscle noise, instrumentation noise, etc. Our proposed solution surpasses these problems by automatically selecting the optimal region for a further post-processing stage, according to a criteria defined by the user. In our second experiment, the effect of the  optimal window is measured independently if cardiac transition constraints are satisfied or not. To do so, the optimal window is obtained by solving the linear program equation \eqref{Shortest_path_ExtraConstraints}. But instead of returning the corresponding state output variables, it returns the sequence generated by applying an argmax function, column-by-column to the state output probability distribution, as it is explained in Section \ref{Standard_Approaches}. The results are displayed in Figure \ref{fig:MC7} in solid lines. As it is possible to observe, the results are not different when compared to our first experiment. The same aforementioned rationale applies in this case, inside of this interest window, both algorithms are confident in assigning a state to a sample, and perhaps it is inside of this window, where state transition constraints are easily obeyed by both methods. 


\section{Conclusion}\label{sec:Conclusions}

In this paper, a novel algorithm capable of selecting the optimal region of interest is proposed. The proposed new feature, enables the user to automatically select the ideal window, where the next post-processing stage is going to happen, according to a set of criterions defined by user itself. As a result, it is very likely that more robust target features are going to be selected, thus consequently increasing the robustness of the system, e.g in detecting abnormalities in physiological signals. Furthermore, the algorithm accurately predicts the hidden state sequence of events in physiological signals, thus making it perhaps more acceptable to physicians. 
For future work, we are going to study an optimal criteria to select the most unusual sound waves, e.g a large S2 split, which could be an indicator of pulmonary hypertension. 




\section*{Acknowledgment}

This work is a result of the Project DigiScope2 (POCI-01-0145-FEDER-029200 - PTDC/CCI-COM/29200/2017), funded by Fundo Europeu de Desenvolvimento Regional (FEDER), through Programa Operacional Competitividade e Internacionalização (POCI), by national funds, through Funda\c{c}\~ao para a Ci\^encia e Tecnologia (FCT), and partially financed by Institut de valorisation des donn\'ees and Fonds de Recherche du Qu\'ebec, through the FRQ-IVADO Research Chair in Data Science for Combinatorial Game Theory.

\ifCLASSOPTIONcaptionsoff
  \newpage
\fi


\bibliographystyle{IEEEtran}
\bibliography{ref}

\end{document}